\documentclass[%
aip,
amsmath,amssymb,
reprint%
]{revtex4-1}

\usepackage{dblfloatfix}
\usepackage{graphicx}
 
\usepackage{amsmath}
\usepackage{mathtools}
\usepackage[titletoc, toc, page]{appendix}
\usepackage{chemformula}
\usepackage{amssymb}
\usepackage{float}
\usepackage{subfig}
\usepackage{afterpage}%
\usepackage{subfig}
\usepackage{caption}
 \usepackage{etoolbox}

\begin{document}
	\author{Shiqi Zhou}
	\affiliation{	School of Physics and Electronics, Central South University, Changsha, 410083, Hunan, China}
	
	\author{Amin Bakhshandeh}
	\email{bakhshandeh.amin@gmail.com}
	\affiliation{Instituto de F\'isica, Universidade Federal do Rio Grande do Sul, Caixa Postal 15051, CEP 91501-970, Porto Alegre, RS, Brazil.}

	\title{{\color{black}Interaction   between two overall neutral charged microscopically patterned surfaces} }

	\begin{abstract}

{\color{black}We study the interaction between heterogeneously charged surfaces in an electrolyte solution by employing classical Density Functional Theory (cDFT) and Monte Carlo simulations. We observe a consistent behavior between cDFT and Monte Carlo simulations regarding force curves and two-dimensional density profiles. Armed with the validated cDFT, we explore the system’s behavior under parameters challenging to simulate directly. Our findings include impacts of domain size, domain charge, domain charge configuration, and bulk electrolyte concentration on the osmotic pressure. Remarkably,  the force curve is more sensitive to the domain size  for  asymmetric configuration than symmetry configuration; the bulk concentration weakly influences the force curve independent of  the system configurations.} 
	\end{abstract}
	\maketitle
	\section{ Introduction}
Electrostatic interactions play a critical role in stabilizing colloidal systems, ionic chemical reactions, and biochemical and physical phenomena~\cite{butler2003ion,vsamaj2011counterions,levin2002electrostatic,fisher1993criticality,ohshima2018approximate,caetano2017critical,de2014critical,de2015inverted,2018195,C9SM02532D}. These interactions are responsible for many exciting phenomena; as a result, their study has led to significant advances in various scientific fields.
	
	One intriguing phenomenon observed in colloidal suspensions with multivalent ions is the reversal of electrophoretic mobility~\cite{levin2002electrostatic}. Under certain conditions, a like-charge attraction between colloidal particles of the same charge sign can also occur ~\cite{levin2002electrostatic}.
	Understanding electrostatic interactions and their effects on colloidal systems is crucial for designing novel materials and processes in various applications, such as drug delivery, energy storage, and water treatment. Therefore, it is important to continue investigating the mechanisms and properties of electrostatic interactions in various systems.

	When a charged surface is in contact with an electrolyte solution, an electrical potential difference is created across the interface, which attracts oppositely charged ions and creates an electric double layer (EDL). Helmholtz made the first attempt to formulate the concept of EDL in 1853~\cite{helmholtz1853ueber}, who proposed a primitive model in which there is a boundary between the charged surface and the electrolyte solution that constitutes an electrical double layer, with a layer of oppositely charged ions at the surface. However, Helmholtz's model did not consider the effect of the thermal motion of ions. This deficiency was corrected by Gouy-Chapman (GC) ~\cite{gouy1910constitution,chapman1913li}, who introduced the concept of a diffuse double layer.
	
	The stability of colloidal particles is more complicated and cannot be explained by the GC model alone. To provide a general understanding of the stability of colloidal suspensions, there are some approximate theories, such as Debye–Hückel and Derjaguin–Landau–Verwey–Overbeek (DLVO) theory. However, the problem with these theories is that they have limitations and can only be applied to systems such as 1:1 electrolytes with moderate concentrations. As the correlation of the system increases, these theories become inadequate and fail ~\cite{derjaguin1941theory,verwey1948theory,zhou2017effective}.
	
	Recent studies have shown that the interaction between the charged surface and the surrounding biomolecules can heavily influence the behavior of charged surfaces in biological systems. In particular, the presence of charged lipids and other membrane components can significantly alter the behavior of charged surfaces on cell membranes, leading to new and exciting phenomena~\cite{o2003intermolecular,mm}, recent developments in experimental techniques, such as single-molecule force spectroscopy, have provided new insights into the behavior of charged surfaces in biological systems~\cite{sumbul2019single,zlatanova2000single}.
	
	Recently, research has focused on the long-range interactions between heterogeneously charged surfaces. These surfaces are significant in nanotechnology, as it is possible to create periodically charged patterned surfaces using nano-fabrication techniques~\cite{parthasarathy2005electrostatically,sayin2017formation}. These systems exhibit unique behavior, such as the adsorption of polyelectrolytes and polyampholytes being dependent on their configuration~\cite{bakhshandeh2019adsorption,D0SM01386B,C8SM00226F,drelich2011charge,velichko2006electrostatic,bakhshandeh2019adsorption,b11,bakhshandeh2019adsorption,bakhshandeh2015interaction}. It has been observed experimentally that the attraction between these surfaces can extend up to 500~\AA~\cite{silbert2012long}. One possible explanation for such observation is the correlation between the charged domains on the surfaces~\cite{bakhshandeh2015interaction}. If this assumption was correct, with the rapid shear movement of plates, the attraction would disappear; in this situation, the domains do not have time to adjust themselves, and as a result, the correlations and attraction force get weakened. However, it is known that the correlation does not play any role in this observation, and the attraction force is due to pure electrostatic interaction~\cite{bakhshandeh2015interaction,silbert2012long}. {\color{black}In general, the study of these systems can be complex and often requires the use of molecular simulations and advanced approaches such as classical Density Functional Theory (cDFT)} \cite{Roth_2010,mussotter2020heterogeneous,zhou2022effective,zhou2019effective,zhou2017classical}.  
	 
The classical density functional theory provides a powerful tool for the calculation of the
structure and thermodynamic properties of heterogeneous fluid systems~\cite{zhou2020statistical,lutsko2010recent,monson2012understanding,lowen2002density,zhou2000density,zhou2017new,mussotter2018electrolyte,zhou2020effective,zhou2020inter,zhou2020effective}.  Most studies validating the accuracy of cDFT are primarily focused on one-dimensional cases, where the density distribution is solely dependent on a one-dimensional coordinate. {\color{black} Moreover, for three-dimensional cases, various scenarios have been compared with molecular simulations, resulting in different reported outcomes~\cite{hartel,schoonen,lam2018solvent}.}

 {\color{black}	 The effectiveness of cDFT in the present two-dimensional model is still unknown, despite its demonstrated utility in studying complex systems, such as heterogeneously charged surfaces within electrolyte solutions ~\cite{zhou2022effective,zhou2020effective,zhou2023effective,zhou2019effective,zhou2020inter,zhou2017effective,zhou2020statistical,zhou2020effective}. Therefore, one of the aim of this article is to assess the performance of a commonly used cDFT version in a two-dimensional case using molecular simulation data, since to best of our knowledge such comparison does not exist.   }

	Despite the advances that have been made in our understanding of charged surfaces in electrolyte solutions, there is still much to be learned. The behavior of these systems is highly dependent on the properties of the electrolyte solution, such as its concentration and ionic strength. Furthermore, the effects of thermal fluctuations and the presence of other molecules in the system can also significantly impact the behavior of charged surfaces. Additional research is required to fully comprehend the behavior of these complex systems and to create new theoretical and computational tools that can help us better predict their behavior in different environments.
	
The behavior of these systems is governed by two forces: electrostatic and entropic. The entropic force in an electrolyte solution arises from the collisions between an ion with other ions or surfaces within the solution., resulting in an exclusion volume effect. This effect causes the particles to experience a net force that is proportional to the density gradient of the surrounding particles.

This paper presents a comprehensive study of the interaction between two heterogeneously charged neutral surfaces (HCNS) using molecular simulation and cDFT.  Through our simulations and theoretical calculations using cDFT, we aim to shed light on the complex interplay between electrostatic interactions and entropic forces that govern the behavior of these systems.  
	
	The paper is organized as follows: In section~\ref{s1}, we provide a detailed description of our simulation approach and theoretical model used to investigate the behavior of heterogeneously charged neutral surfaces. In section~\ref{s2}, {\color{black}we describe the cDFT,} In section~\ref{s3}, we present and discuss our results; in section~\ref{s4}, we conclude our work.
	\section{Simulation method}~\label{s1}
\begin{figure}[H]
		\centering
		\includegraphics[width=0.7\linewidth]{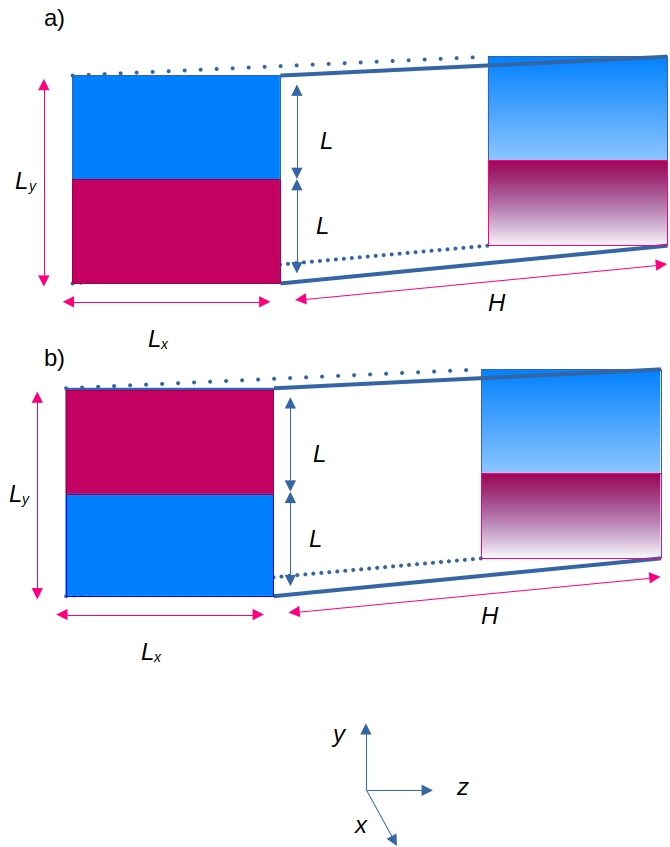}
		\caption{The schematic representation of the simulation cell used in this study. (a) symmetric configuration. (b) asymmetric configuration. There are periodic boundary conditions along the $x$ and $y$ coordinates.}
		\label{fa}
	\end{figure}

{\color{black}We utilize a two-dimensional model to investigate} the behavior of heterogeneously charged surfaces immersed in an electrolyte solution. The model consists of two flat surfaces with dimensions $L_x$ and $L_y$, separated by a distance of $H$ and surrounded by the electrolyte. For our simulations, we set $L_x=L_y=400$ Å. The solvent is treated as a uniform dielectric with permittivity $\epsilon_w$ and   Bjerrum length   is defined as $\lambda_B = q/k_B T \epsilon_w$, where $q$, $k_B$, and $T$ denote the proton charge, Boltzmann constant, and temperature, respectively. We take $\lambda_B$ to be 7.2 Å. Each plate has a charged domain with dimensions of $L \times L_x$. The cell configuration is shown in Fig.~\ref{fa}.
	
	Our model considers ions as hard spheres with a radius of 2~\AA, and we simulate  the system using the grand canonical Monte Carlo (GCMC) algorithm, as described in previous studies~\cite{valleau1980primitive,frenkel1996understanding}. To study the effect of the electrolyte solution, we put the system in contact with a salt reservoir at a concentration of $c$, and determine the excess chemical potential of the reservoir using the mean spherical approximation (MSA)~\cite{ho2003interfacial,waisman1973radial,waisman1972mean,blum1975mean}. {\color{black}Although the MSA is an approximation, it is very precise for 1:1 electrolytes ~\cite{bakhshandeh2022widom}}. 
	
	In this paper, we use the term "\textit{symmetric}" when two plates have the same configuration, and "\textit{asymmetric}" {\color{black}when the configurations of two plates are mutually crosswise arranged}. 

\begin{figure}[H]
		\centering
		\includegraphics[width=1.1\linewidth]{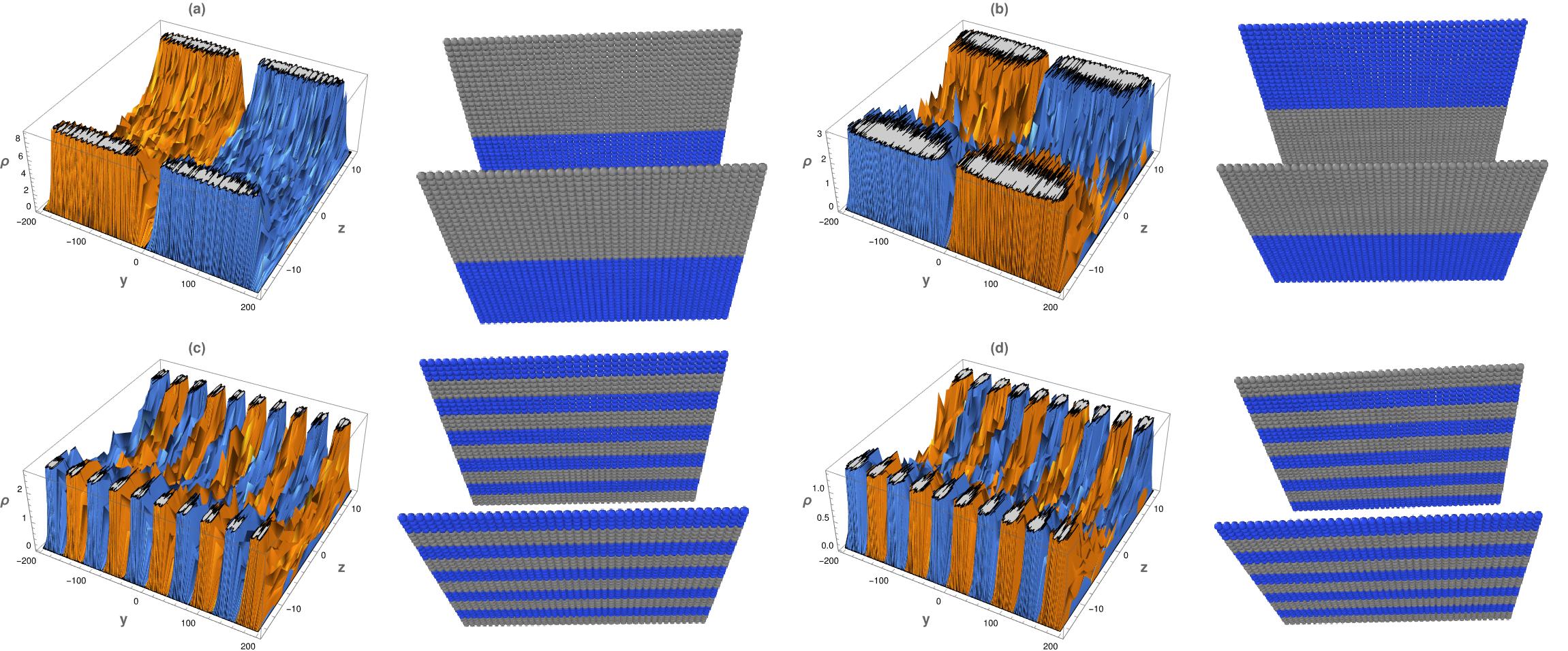}
		\caption{The density profile of negative and positive ions for different patterns. The  surface charge density of the plates is 0.0998 Cm$^{-2}$, and the concentration of salt is 20mM. The plate configuration is shown next to the density profile for each case. The figure utilizes a distinct color code to represent different charges.  
		 }
		\label{fig:11}
	\end{figure}

	We implemented a grand canonical Monte Carlo algorithm (GCMC) simulation for salt concentration 20 and 100 mM for surface charge density 0.0998 and 0.2 Cm$^{-2}$. Each plate consists of 1600 charged particles arranged in different patterns. To evaluate the electrostatic energy, we used the 3D Ewald summation method with a correction for the slab geometry of Yeh and Berkowitz~\cite{yeh1999ewald}.
	In Fig.~\ref{fig:11}, we plotted the density profile of negative and positive ions in 3D for different patterns.
	

	To evaluate the force on the plate, we consider both the electrostatic interactions and the entropic force resulting from the momentum transfer of ions colliding with the plates. To calculate the entropic force, we use the method proposed by Wu \textit{et al}.~\cite{wu1999monte,bakhshandeh2015interaction}:
	
	For each sample, we move the plate towards the other plate and count the number of overlaps with electrolyte ions~\cite{ bakhshandeh2015interaction}:
	\begin{equation}
		\beta F = \frac{\left< N \right>}{\Delta z} \ ,
	\end{equation}
	where $N$ is the number of overlaps of ions and  $\Delta z$ is the displacement of the wall where $\Delta z \rightarrow 0$. The entropic pressure is then becomes:
	\begin{equation}
		\beta P = \frac{\left< N  \right>}{\Delta z L_x L_y} \ .
	\end{equation}

	\section{Classical Density Functional Theory}~\label{s2}
	
{\color{black}We use a cDFT version~\cite{zhou2017classical} that has been repeatedly validated in several one-dimensional cases. In this version, the treatment of hard sphere repulsive interactions incorporates a recently developed extended form of Ref.~\cite{zhou2011enhanced} of the fundamental measure functional version proposed by Kierlik and Rosinberg~\cite{kierlik1990free}  }.	
	   The long-range Coulomb interaction is treated with a mean field approximation, and coupling between hard sphere and Coulomb
	interactions is treated with second-order perturbation expansion based on mean
	spherical approximation. Although this version has been repeatedly validated in
	one-dimensional cases, two-dimensional validation has not yet been reported.
	Consideration of two-dimensional model increases difficulty in algorithm
	convergence. For this reason, we take three measures. (i): the calculation is
	started from zero surface charge density; one gradually increases the charge
	density to the target value and takes the density distribution output of the
	previous charge density as input for the next charge density calculation. (ii) The
	nonlinear equations resulting from the discretized two-dimensional cDFT
	equation are solved by the Newton GMRES algorithm which is implemented in
	the public-domain nonlinear Krylov solver NITSOL. (iii) {\color{black}In one-dimensional cases, the mesh size is typically very small, often around 0.025 times the molecular diameter. However, in the two-dimensional case, we set the mesh size to 0.1 times the molecular diameter for both dimensions. We conducted a validation test comparing grid sizes of 0.05 and 0.1 times the molecular diameter and found nearly identical results.}

{\color{black}To evaluate the osmotic pressure, we begin by calculating the effective electrostatic interaction potential as a function of the plate separation $H$. This potential is obtained as the difference between the excess grand potential when the two plates are at a distance of $H$ apart and the excess grand potential when the plates are sufficiently far apart. Next, the derivative of this interaction potential with respect to distance is computed to obtain the interaction force between the plates.
	
	The excess grand potential is determined by subtracting the grand potential of the system with the two plates at a distance of $H$ apart by the grand potential of a bulk system with the same volume but without the two plates. Within the framework of cDFT, it is possible to calculate the grand potential of a heterogeneous system, such as the two-plate system, by substituting the density distribution obtained from cDFT into the expression for the grand potential. Conversely, if the bulk density is used as input, the grand potential of the bulk system can be obtained.
	
  }

	\section{Results \& Discussion}~\label{s3}
\begin{figure}[H]
		\centering
		\includegraphics[width=1.1\linewidth]{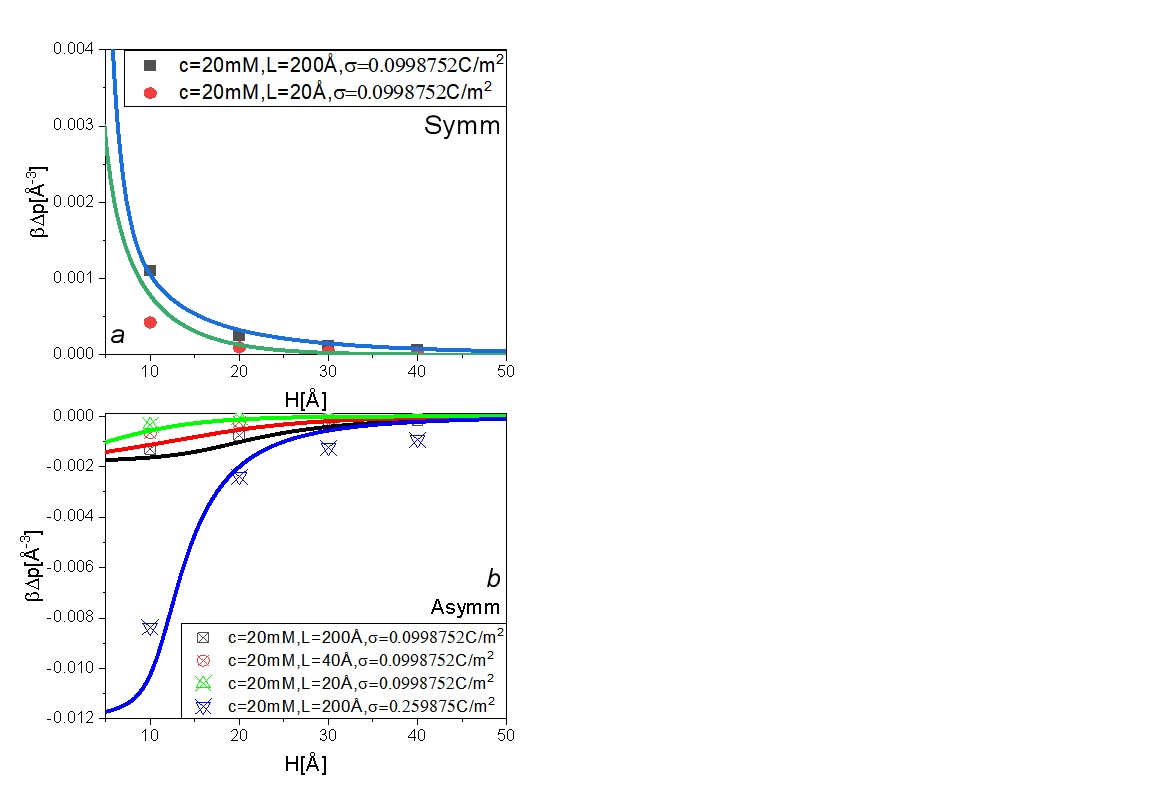}
		\caption{Osmotic pressure {\color{black}	against separation distance between plates, $H$,} for two different surface charge densities and domain size for symmetric and asymmetric configuration of plates . The solid lines are cDFT, and the symbols result from MC simulations. (a)  Symmetric configuration. (b)Asymmetric configuration. }
		\label{fig2}
	\end{figure}

 In the first step, we compared the forces obtained by Monte Carlo (MC) simulations and cDFT for  symmetric and asymmetric cases with a domain size of 200 $\times$ 400\AA, and a surface charge density of 0.0998 Cm$^{-2}$ in the presence of 20 mM 1:1 electrolytes. As shown in Fig.~\ref{fig2}(a-b) for symmetric and asymmetric cases, there is good agreement between simulation and cDFT.
	
	
\begin{figure}[H]
		\centering
		\includegraphics[width=1.1\linewidth]{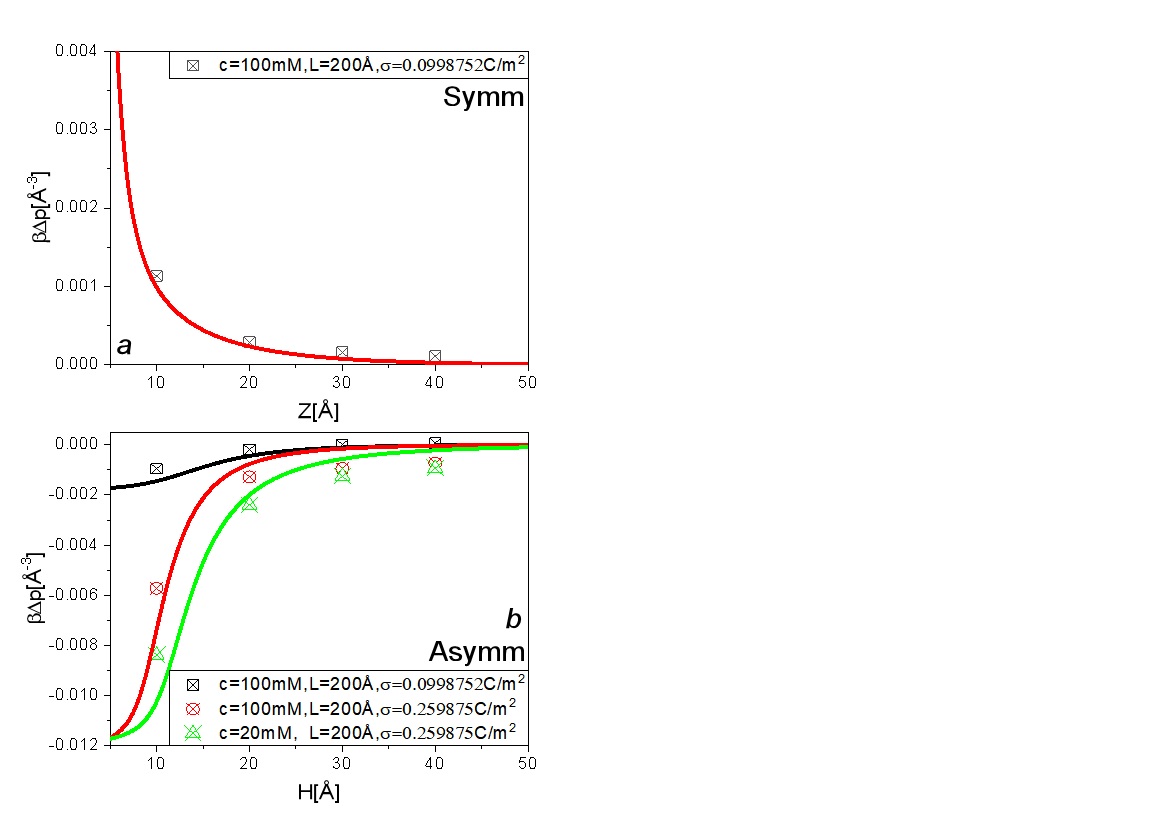}
		\caption{Osmotic pressure {\color{black}	against separation distance between plates, $H$,} for different surface charge densities and domain size for symmetric and asymmetric configurations of plates. The solid lines are cDFT, and the symbols result from MC simulations. }
		\label{fig3}
	\end{figure}

	Next, we investigate the effect of electrolyte concentration on the interaction between the charged microscopically patterned surfaces. For the symmetric case, Fig.~\ref{fig2}(a) with a domain size, 200 $\times$ 400~\AA~ and a surface charge density of 0.0998 Cm$^{-2}$, the repulsion force does not change significantly when the concentration increases from 20 to 100 mM. The same observation can be seen for the asymmetric case where the attraction exists. However, this is not the case for surface charge density. As is seen in Fig.~\ref{fig2}(b), the blue curve representing the case with a surface charge density of 0.25 Cm$^{-2}$ exhibits a significant change in the attraction curve. In Fig.~\ref{fig3}(a), the osmotic pressure is plotted against the separation distance. As can be seen, despite increasing the electrolyte concentration by five times, the osmotic pressure for the symmetric case does not change significantly.
	
\begin{figure}[H]
		\centering
		\includegraphics[width=1.1\linewidth]{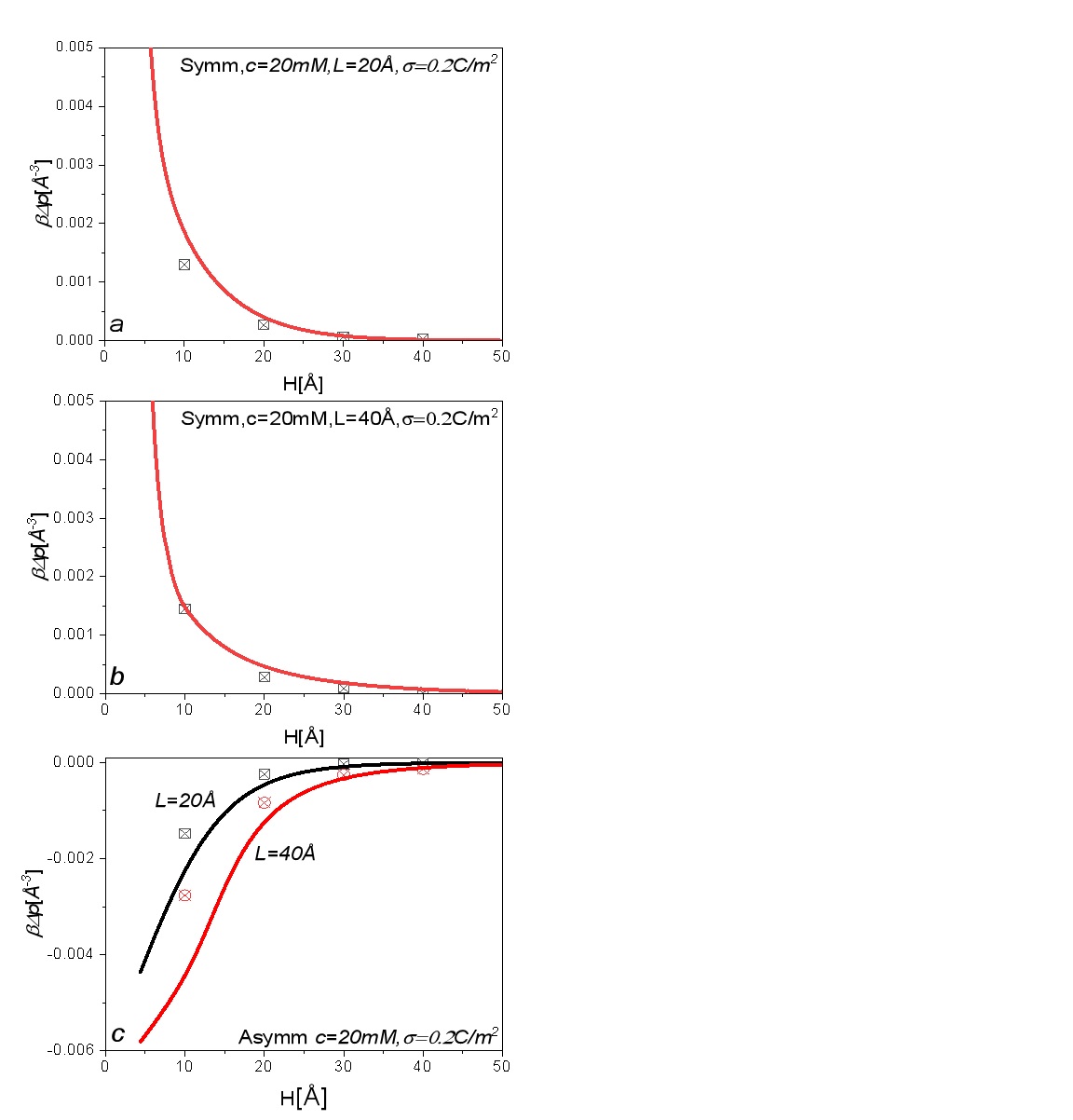}
		\caption{Osmotic pressure  {\color{black}	against separation distance between plates, $H$,} for different surface charge density and domain's size for symmetric and asymmetric configuration of plates. The solid lines are cDFT and symbols are result of MC simulations.}
		\label{fig4}
	\end{figure}

\begin{figure}[H]
		\centering
		\includegraphics[width=1.\linewidth]{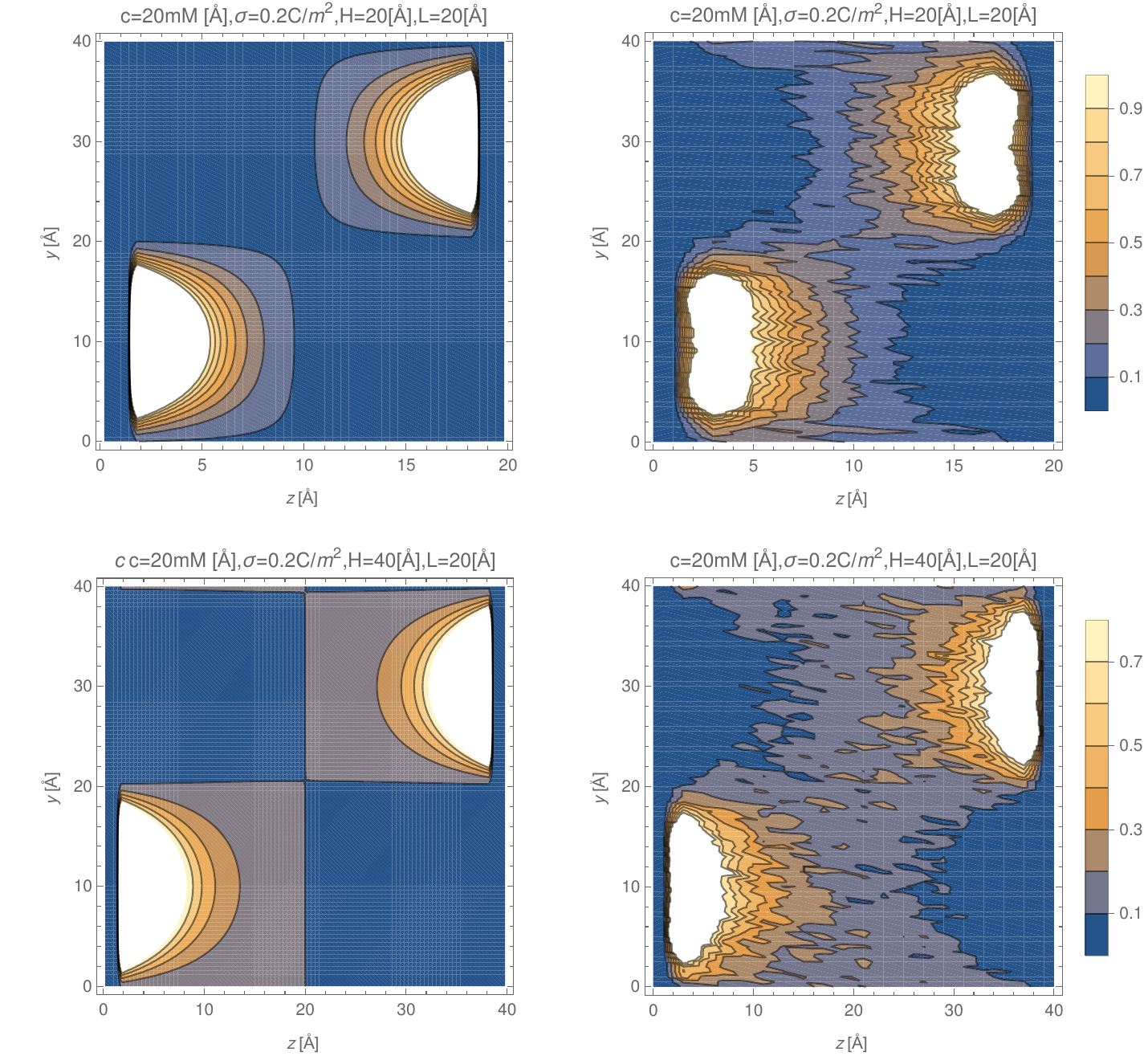}
		\caption{2D density profile of ions for simulation and cDFT for asymmetric configuration. The separation distance for the first row and second raw is  $H=20$ and 40~\AA~  respectively. The concentration of 1:1 electrolyte is 20mM  and the surface charge density is 20Cm$^{-2}$.  The domain's size is 20$\times$400~\AA$^2$. The first column is density predicted by cDFT, and the second column results from MC simulation. Concentration is in M.}
		\label{fig5}
	\end{figure}

\begin{figure}[H]
		\centering
		\includegraphics[width=1.\linewidth]{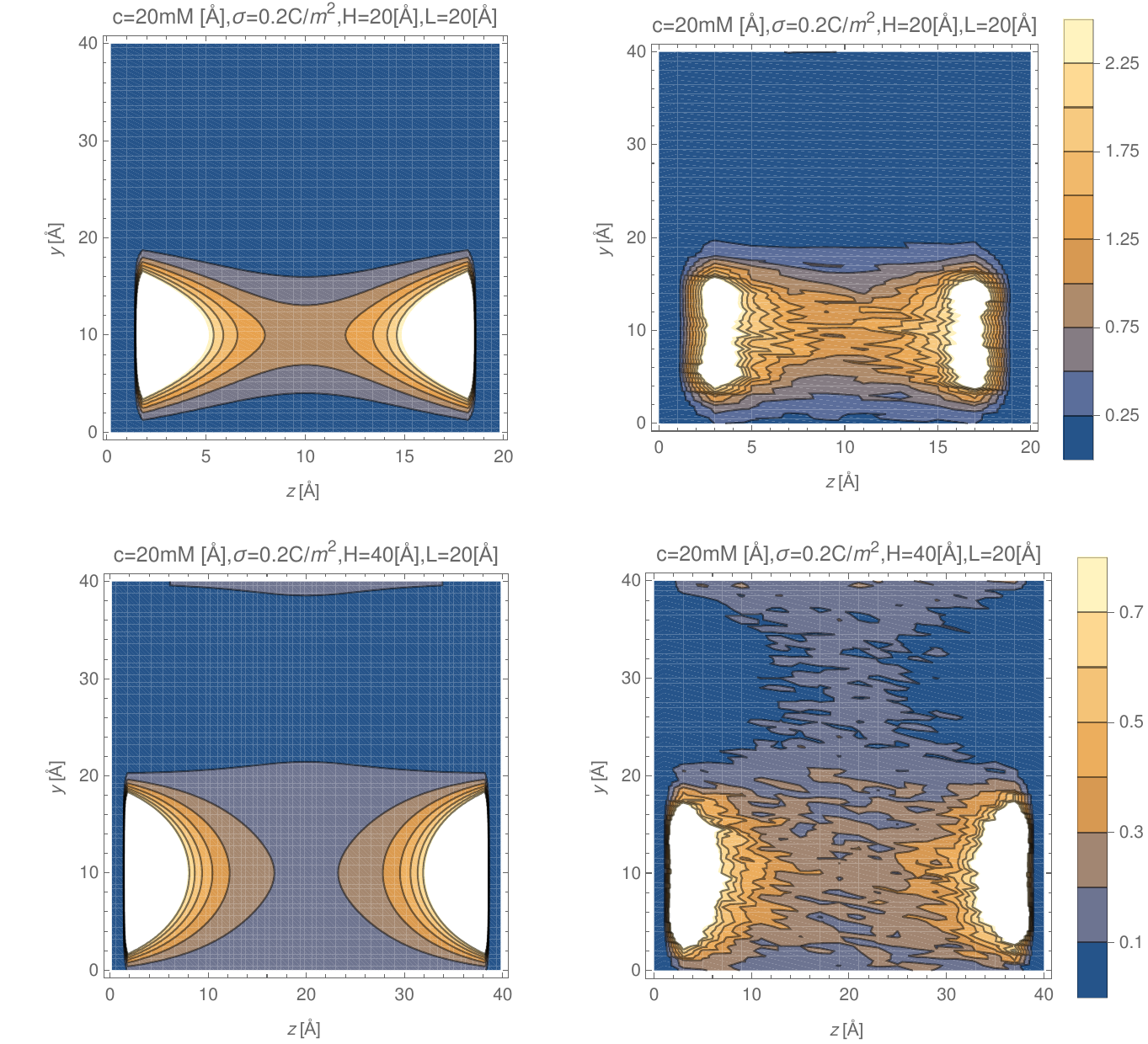}
		\caption{2D density profile of ions for simulation and cDFT for symmetric configuration. The separation distance for the first row and second raw is  $H=20$ and 40~\AA~  respectively. The concentration of 1:1 electrolyte is 20mM  and the surface charge density is 20Cm$^{-2}$.  The domain's size is 20$\times$400~\AA$^2$. The first column is density predicted by cDFT, and the second column results from MC simulation. Concentration is in M.}
		\label{fig6}
	\end{figure}

\begin{figure}[H]
		\centering
		\includegraphics[width=1.0\linewidth]{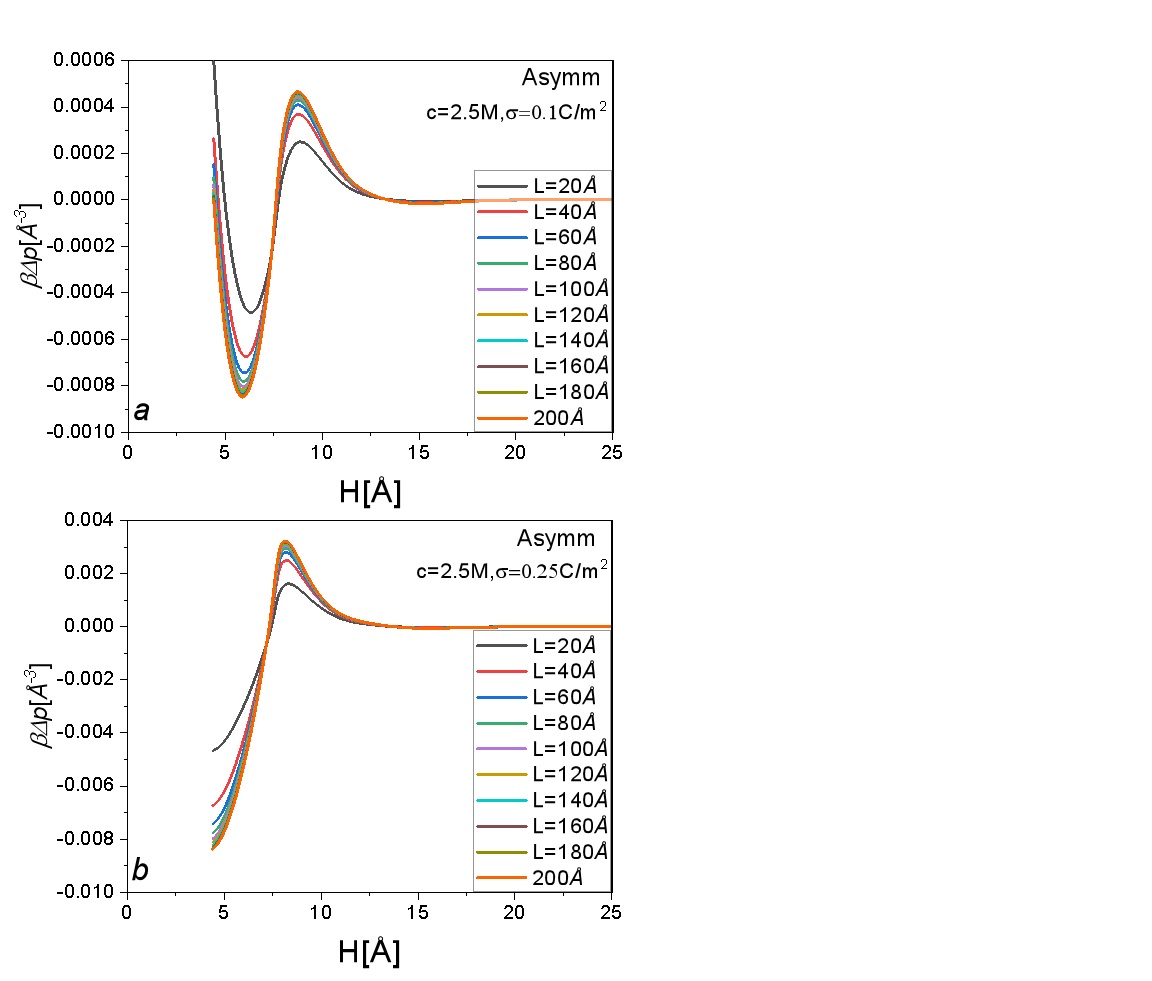}
		\caption{{\color{black}Osmotic pressure for asymmetric case for different domain sizes. (a) $\sigma=$0.1 Cm$^{-2}$, (b)~$\sigma=$ 0.25 Cm$^{-2}$, the concentration of electrolyte is fixed at 2.5 M.}}
		\label{fig7}
	\end{figure}

	However, as seen in Fig.~\ref{fig3}(b), the attraction is higher for 20 mM electrolyte concentration than 100 mM. The observed phenomenon can be attributed to the entropic forces, wherein the increased collision between the plates and ions of the electrolyte at higher concentrations leads to a reduction in the attractive force. However, as the two plates get closer, the entropic forces reduce, since the number of particles between plates reduces, for both cases  resulting in the same attraction force magnitude.  
As is shown in Fig~\ref{fig4}(a-b), for the symmetric case with a surface charge density of 0.2 Cm$^{-2}$, the repulsion does not change significantly for different domain sizes of 20$\times$200 \AA$^2$ and 40$\times$200 \AA$^2$. However, there seems to
be greater sensitivity to domain size for asymmetric cases in which there is the attractive force. It is observed that a bigger domain size results in a significantly higher attraction between plates for asymmetric configurations.
	Also, we compared the density profiles of ions obtained from Monte Carlo simulation and cDFT in Figs.~\ref{fig5} and~\ref{fig6}, to achieve this, we mapped the 3D density profile of ions to 2D. As is observed, the MC simulation results have higher fluctuations than cDFT, but both simulation and cDFT are consistent overall. The ion density profiles provide us with insight into the structure of the electric double layer around the charged patterned surfaces.  {\color{black}The issue with the density profile is that its fluctuations are unavoidably large, as depicted in Figs.~\ref{fig5}-\ref{fig6}, this is due to the fact that to obtain 2D density profile, the small bins were used and this led to high fluctuations. However, overall, the agreement is generally acceptable. It is worth noting that for the two-dimensional system, the force curves do not appear to be significantly influenced by the density profile. Although the density profile is a physically significant quantity that sensitively depends on the microscopic configuration, its impact on the force curves seems to be relatively minimal. This observation is satisfactorily predicted by cDFT. Since force is the
		negative derivative of the potential function with respect to distance, and its effective prediction should depend more sensitively on the accuracy of the method. Fortunately,  cDFT demonstrates promising accuracy in predicting force curves. However, this particular issue is the subject of our future research.} 
	After comparing the predictions of cDFT and Monte Carlo simulations, we observed a good agreement between the two methods. This comparison confirms the accuracy and reliability of our theoretical approach to studying the behavior of these systems. As a result, we focused solely on cDFT to investigate systems with higher electrolyte concentrations (2.5 M) and surface charge densities (2.5 Cm$^{-2}$). These parameters are challenging to simulate using MC method due to the difficulty in reaching equilibrium. As is seen in Fig.~\ref{fig7} we plot the osmotic pressure for different domain sizes at different separation distances for asymmetric cases.
	{\color{black} Fig.~\ref{fig7} shows no attraction between the plates at high electrolyte concentrations for $H>10$~\AA. This is opposite to the case of low electrolyte concentration, in which the attraction exists at separation distances up to  40~\AA~ for asymmetrical pattern.
		
	In  Fig.\ref{fig7}(a), we show osmotic pressure obtained from cDFT   for asymmetric configurations of plates with surface charge density $\sigma = 0.1$ Cm$^{-2}$ and different domain sizes in the presence of 2.5M 1:1 electrolyte. As is seen in Fig.\ref{fig7}(a) there is a maximum attractions and repulsion forces which around 5.9 and 8.8~\AA. When the separation distance between plates is small, the higher entropic forces encourage ions to migrate from the region between the plates to the reservoir. Consequently, this leads to an increased Debye length and attraction force between the plates. However, by increasing the separation distance, $H$, the Helmholtz double layer (HDL) starts to form. The second maximum  occur at a distance approximately twice the ion diameter (8 \AA). This distance corresponds to when the Helmholtz layers of the two plates come into contact with each other. This double layer, increase the excluded volume between plates and as a result increases the entropic force, this results in the plates do not experiencing any attraction and only an entropic force being observed. When the separation distance increases, the entropic force gets smaller, and also, since the Debye length is around 1.9~\AA, the interaction force rapidly goes to zero.

	In Fig.\ref{fig7}(b), as the surface charge density increases, the first minimum (maximum attraction force) disappears due to the direct electrostatic interaction between the plates. However, the maximum repulsion force shifts to a smaller separation distance of around 8.1~\AA, which can be attributed to the complete formation of the HDL on the domains. This is because the higher surface charge density leads to a stronger binding between ions and plates,and in turn this leads to   the formation of a more rigid HDL.}

\begin{figure}[H]
		\centering
		\includegraphics[width=1.0\linewidth]{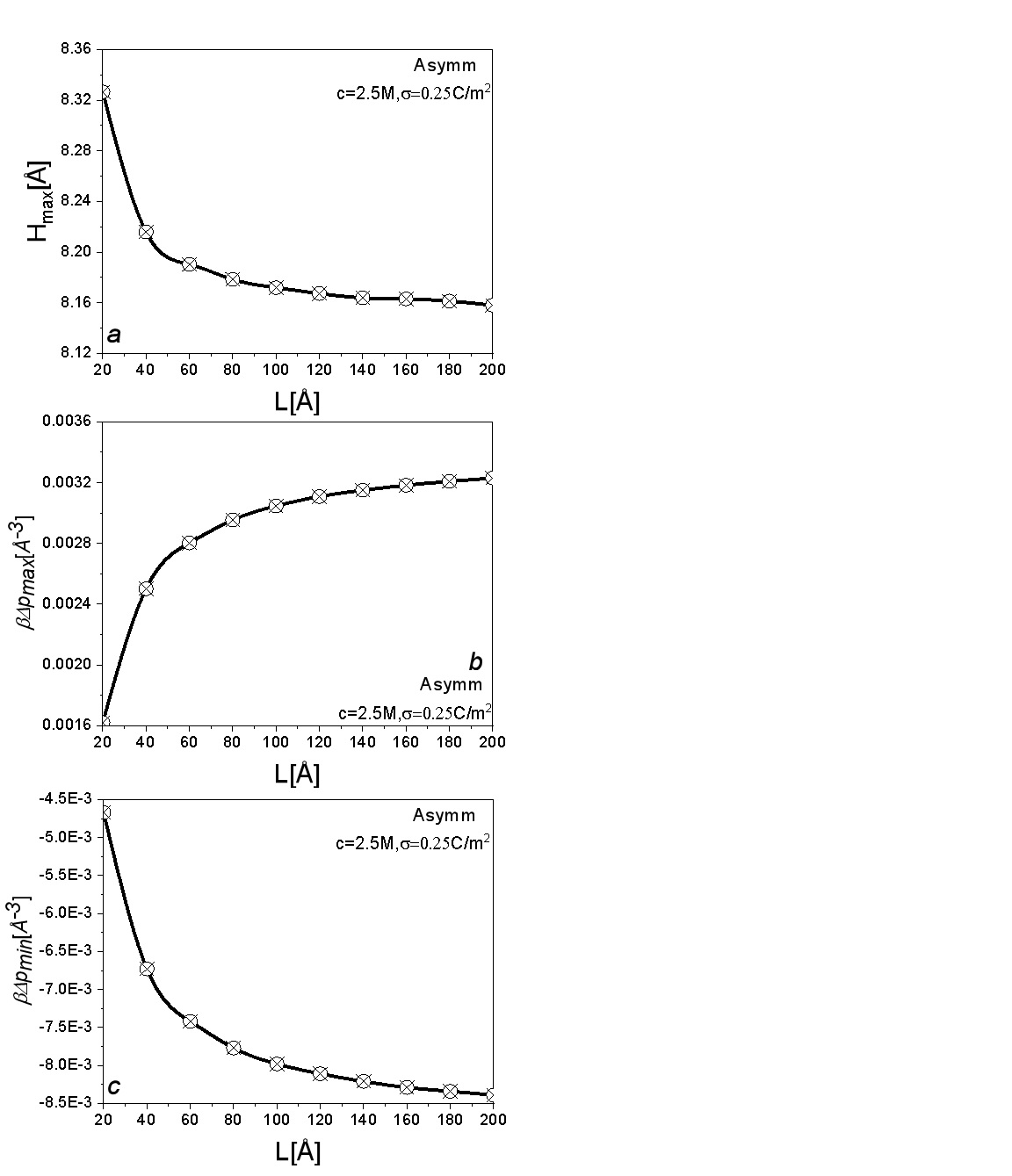}
		\caption{(a) Represents the separation distance at which maximum repulsion force    occurs. (b)  Maximum force observed  in Fig.~\ref{fig7} against different domain's sizes. (c) Minimum force observed  in Fig.~\ref{fig7} against different domain's sizes. $\sigma=0.25$ Cm$^{-2}$.  The lines are provided as a visual aid.}
		\label{fig8}
	\end{figure}

\begin{figure}[H]
		\centering
		\includegraphics[width=0.7\linewidth]{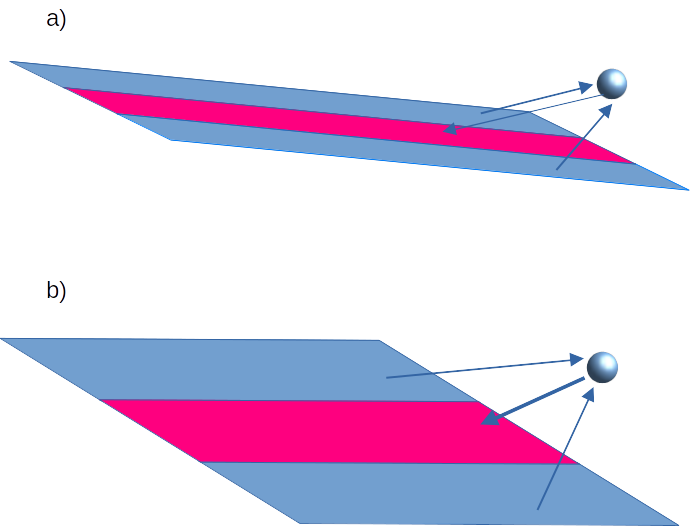}
		\caption{Schematic representation of ion approximation to domains with opposite charge. (a) When an ion approaches a small domain with opposite charge, it experiences a high repulsion force from neighboring domains. (b) When an ion approaches a larger domain with opposite charge, it experiences a reduced repulsion force from neighboring domains.}
		\label{fl}
	\end{figure}

	In Fig.~\ref{fig8} we plot the maximum and minimum forces observed in Fig.~\ref{fig7} against different domain's size and the $H$ which the maximum repulsion force appears in terms of $L$.
	As is shown in Fig.\ref{fig8}(a), the maximum repulsion occurs at smaller distances for larger domain sizes. This suggests that HDL forms easily and completely for larger
	domains than smaller ones. This observation is because the ions find it more difficult to approximate small domains due to the repulsion from neighboring domains, we show this schematically in Fig.~\ref{fl}. The maximum osmotic pressure, Fig.\ref{fig8}(b), reveals that as the domain size decreases, the maximum force also reduces. This, again, can be explained by the fact that the ions are firmly anchored to the charged surface within the HDL of the larger domains, where ions can more easily approximate the domain due to the weaker repulsion of neighboring domains. Additionally, as explained previously,  Fig.\ref{fig8}(c) the maximum attraction force directly relates to the domain size. This is due to the electrostatic interaction between domains, which overcomes the entropic forces. 
	
\begin{figure}[H]
		\centering
		\includegraphics[width=1.0\linewidth]{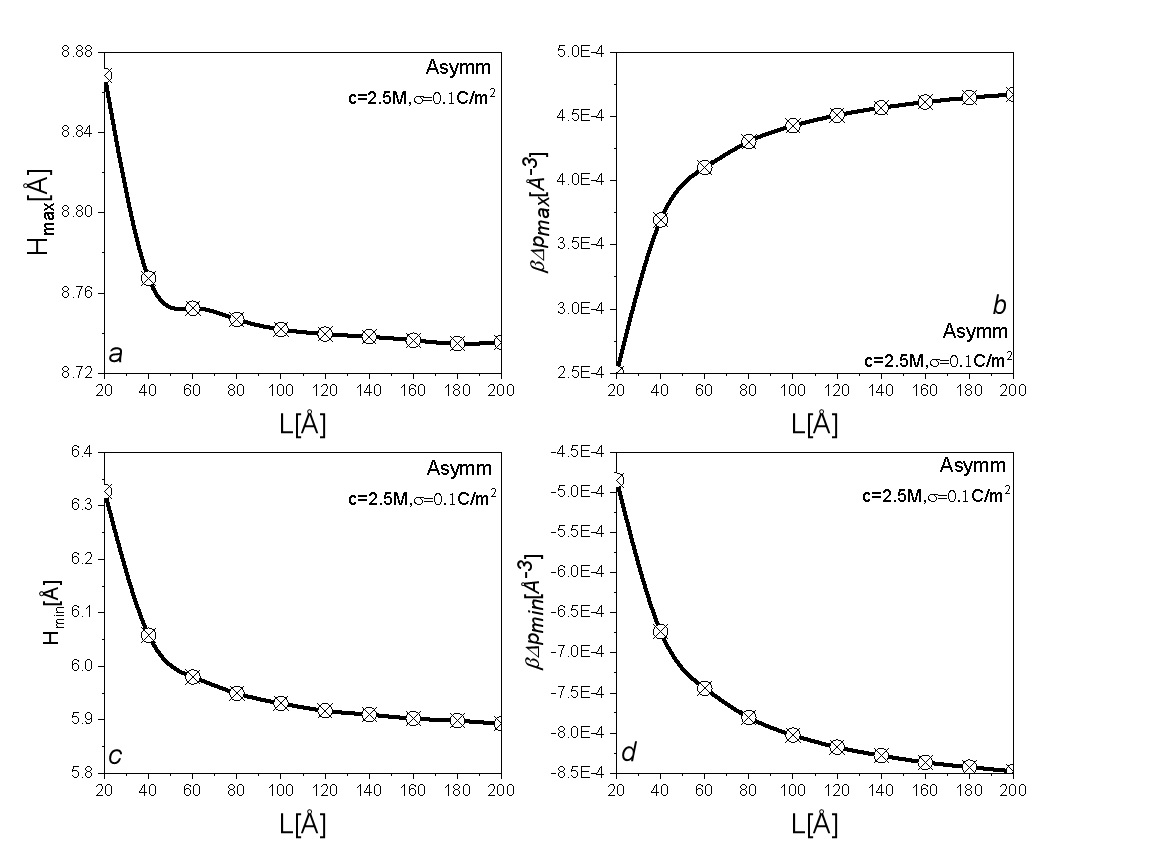}
		\caption{(a and c) {\color{black}Separation distances (for maximum and minimum force, respectively, as observed in  Fig.~\ref{fig7} against domain size for $\sigma=0.1$ Cm$^{-2}$.  (b and d) Maximum force and Minimum force against domain size for $\sigma=0.1$ Cm$^{-2}$. The lines are provided as a visual aid.}}
		\label{fig9}
	\end{figure}

In  Fig.~\ref{fig9}, the same information is provided for  $\sigma=0.1$ Cm$^{-2}$, and besides that, we also show the separation distance when maximum attraction force occur, $H_{min}$, in terms of $L$ in  Fig.~\ref{fig8}(a).
	As is shown in Fig. ~\ref{fig9}(a and c), the separation distance at which maximum and minimum forces occur exhibits the same trend as Fig. ~\ref{fig8}. For maximum forces in Fig. ~\ref{fig9}(b), a similar trend to Fig. ~\ref{fig8} is also observed. However, in Fig. ~\ref{fig9}(d), it can be seen that a larger domain size leads to stronger attraction forces.
	
	\section{Conclusion}~\label{s4}
{\color{black}This study investigates the impact of domain size, domain charge, domain surface configuration, and bulk electrolyte concentration on the osmotic pressure of charged systems using Monte Carlo simulation and Classical Density Functional Theory (cDFT). 

We examined the interaction force    between plates with symmetric and asymmetric configurations and its relation with domain size. To this end, we studied surface charge densities of 0.0998 and 0.25 Cm$^{-2}$ and a 20mM 1:1 electrolyte for both asymmetric and symmetric configurations with domain sizes of 20, 40, and 200~\AA. In all cases, we observed attraction between the plates in the asymmetric configurations, while repulsion was observed in the symmetric configurations. Furthermore, we compared the results obtained from recently developed cDFT for plates with non-homogeneous charge distribution with those from MC simulation and found a good agreement.

 Our findings reveal that the domain size has minimal effect on the  osmotic pressure in symmetric configurations. However, the
 attraction between plates is sensitive to the size of the domain for the case of asymmetric configurations.

Furthermore, we analyze the behavior of the force curve with variations in bulk concentration for symmetric and asymmetric configurations. The force curve remains relatively unchanged with changes in bulk concentration, but higher domain charge densities can amplify its sensitivity. Moreover, asymmetric configurations exhibit more complex behavior at higher electrolyte concentrations. It was observed that in this case a maximum repulsion appears which can be explained by HDL.

In addition, our study confirms  the validity of the classical density functional theory cDFT~\cite{zhou2022effective} for slab geometry with non-homogeneous charged distributions. In our future work,  we employ  the method to study the systems with spherical geometry~\cite{lovsdorfer2013symmetry}, these system are of great importance in colloidal science an biological systems  such as viruses.}

	\section{Acknowledgments}
	This project is   supported by the National Natural Science Foundation of China (Grants 22173117), the High Performance Computing Center of Central South University and CAPES.
	
	\bibliography{ref}
\end{document}